\documentclass[
aps,%
12pt,%
final,%
notitlepage,%
oneside,%
onecolumn,%
nobibnotes,%
nofootinbib,
superscriptaddress,%
noshowpacs]%
{revtex4}
\blankaffiliation
\usepackage{epsfig}

\begin{document}
\title{On Fixed-multiplicity Corrections to Correlators \\ 
}
%
\author{\firstname{Peter} \surname{Filip}}
\email[]{fyziflip@savba.sk}
\affiliation{Laboratory of Particle Physics, JINR, Dubna, Russia} 
\address{Institute of Physics, Slovak Academy of Sciences, Bratislava, Slovakia }
\noaffiliation
\begin{abstract}
Correction terms generated in the correlator analysis due to  
multiplicity-dependent observable mean are investigated.
A procedure for subtraction of such terms from calculated
correlator estimates is suggested and the obtained results are discussed.  
\end{abstract}
\maketitle
\section{Introduction.}
Behavior of strongly interacting  matter under extreme conditions attracts 
interest of physicists for a considerably long time. Motivated partially also by
attempts to explain interesting cosmic-ray observations \cite{CosmicRNP2006} 
it has been assumed, that in relativistic collisions of hadrons or nuclei
a short-living system containing a super-dense thermally equilibrated matter 
can be created in laboratory. Intriguing results of experimental groups at CERN and 
BNL show us that such an effort is a challenging task and interpretations
of the obtained experimental information are still vividly discussed.

It has been suggested \cite{ManjSiss} that calculation of the correlator
ratios may allow us to estimate quantitatively a degree of
thermalization of high-multiplicity thermodynamical systems
created in relativistic collisions of hadrons or nuclei.
This suggestion has been met with a considerable interest \cite{Shimanski}
and recently a procedure for the evaluation of higher-order correlators in a
reasonable computer time has been presented \cite{Amelin}.

On experimental side correlator $C^{p_t}_2$ 
has been evaluated using transverse
momenta of charged particles created in relativistic $Au\!+\!\!Au$ 
collisions at RHIC \cite{STAR} and 
centrality dependence of such correlator has
been studied.
However, it seems that calculation of correlators from real experimental data 
requires a more detailed approach. For example, if the observable mean of 
quantity (e.g.\,$\langle p_t\rangle $) used for calculation of correlators
depends on measured multiplicity, correlators 
can be systematically shifted and obtained results biased.

In the following sections we study quantitatively the influence of
multiplicity-dependent observable mean on correlators
and describe a simple procedure allowing one to subtract 
the introduced systematical bias 
from the calculated correlator estimates.

\section{A Simple Correction to $C_2$ Correlator}
Let us start with global 2-particle correlator $C_2^{(n,x)}$ calculated
from measured characteristics $x_i$ (e.g. transverse momentum $p_t$) of observed 
charged particles with multiplicity $n$. For the sample $A$ containing $N_{ev}$ events 
one has (see Eq.(21) in \cite{Amelin} and also Eq.(2) in \cite{STAR})

\begin{equation}
C_2^{(n,x)} = \frac{1}{N_{ev}}\sum_{N_{ev}}
        \sum\!\!\!\!\!\!\!\sum_{\!\!i \neq j}^{\!\![N_{ij}]}
\frac{(x_i - \langle \bar x_G \rangle)(x_j - \langle \bar x_G \rangle)}{N_{ij}}
\label{C2x}
\end{equation}
where $N_{ij}=n!/(n-2)!$ is the number of all particle pairs in a given event with indexes 
$i \neq j$ (symbol $[N_{ij}]$ reminds double-sum) and $\langle \bar x_G \rangle$ is the global mean of particle 
characteristics $x_i$ (e.g. transverse momentum $p_t$) over all events and particles
used for the analysis.

Because correlators are not sensitive \cite{Amelin} to number of particles $n$ 
selected for their calculation  (see Section 3 and Tab.3 in \cite{Amelin}) one can 
select a fixed number of tracks $n<n_k$ for the correlator evaluation
in the sample of events with multiplicities $n_1<n_k<n_2$.

It can be easily shown that if observable mean $\bar x(n_k)$ in 
events with observed multiplicity $n_k$ depends on event multiplicity $n_k$ 
then calculated value 
of correlator $C_2$ for sample $A$ is shifted by the amount $\Delta_2$ 
which depends on the width of multiplicity interval $(n_1,n_2)$ 
and on properties
of multiplicity-dependent observable mean $\bar x(n_k)$.

\begin{equation}
\text{if} \quad \frac{d \bar x(n) }{d n} \neq 0 \quad \quad ; 
         \quad \quad C_2^{\text{\,calc}} = C_2^{\text{\,true}} +\,\, \Delta_2
\end{equation}

In order to demonstrate this  let us divide event sample $A$ into 
subsamples $A1$ and $A2$ 
with 
multiplicities $(n_1,n_{1a})$ and $(n_{1a},n_2)$ and let us assume (see Fig.1)
that observable mean $\bar x(n_k)$ increases with multiplicity $n_k$
in step-like way: 
$\bar x (n_1\le n_k\le n_{1a}) =  \bar x_{A1} $ and 
$\bar x (n_{1a} < n_k \le n_2) =  \bar x_{A2} $.
For correlator $C_2^{(n,x)}$ 
calculated for event sample A one can write
\begin{equation}
C_2^{A=A1+A2} = \frac{1}{N_{ev}}\bigg[ 
\sum_{N_{\!A1}} \sum\!\!\!\!\!\!\!\sum_{\!\!i \neq j}^{\!\![N_{ij}]}
\frac{(x_i - \langle \bar x_G \rangle)(x_j - \langle \bar x_G \rangle)}{N_{ij}}
+
\sum_{N_{\!A2}} \sum\!\!\!\!\!\!\!\sum_{\!\!i \neq j}^{\!\![N_{ij}]}
              \frac{(x_i - \langle \bar x_G \rangle)(x_j - \langle \bar x_G \rangle)}{N_{ij}}
\bigg]
\end{equation}

It is clear that in subsample $A1$ with observable mean $\bar x_{A1}$
the correlator $C_2$ is calculated using an overestimated global mean $\langle \bar x_G \rangle$
which introduces a static shift of global mean
$\Delta \bar x_{A1} = \bar x_{A1} -\langle \bar x_G \rangle$ in the
correlator calculation. 
Since $(x_i - \langle \bar x_G \rangle) = (x_i - \bar x_{A1} + \Delta \bar x_{A1})$
one obtains for events within sub-sample $A1$ (and similarly $A2$):
\begin{equation}
(x_i - \langle \bar x_G \rangle)(x_j - \langle \bar x_G \rangle)=
(x_i - \bar x_{A1})(x_j - \bar x_{A1}) +
 \Delta \bar x_{A1} [ x_i-\bar x_{A1} + x_j- \bar x_{A1}] +(\Delta \bar x_{A1})^2 
\end{equation}
A substitution into Eq.(3) using $\sum_{N_{\!Ak}} \sum_{i=1}^n(x_i-\bar x_{Ak}) = 0$ (for $k$=1,2) gives
\begin{equation}
C_2^{A1+A2} = \frac{N_{\!A1} \cdot C_2^{A1} + N_{\!A2} \cdot C_2^{A2}}{N_{ev}}
              + \frac{N_{\!A1} \cdot(\Delta \bar x_{A1} )^2}{N_{ev}}
              + \frac{N_{\!A2} \cdot(\Delta \bar x_{A2} )^2}{N_{ev}}
\end{equation}
where the correlator for sub-sample $A1$ (and similarly for $A2$) is
\begin{equation}
C_2^{A1} = \frac{1}{N_{\!A1}}\sum_{N_{\!A1}}
\sum\!\!\!\!\!\!\!\sum_{\!\!i \neq j}^{\!\![N_{ij}]}
\frac{(x_i - \bar x_{A1})(x_j - \bar x_{A1})}{N_{ij}} .
\end{equation}

\begin{figure}[h]
\includegraphics[width=12cm]{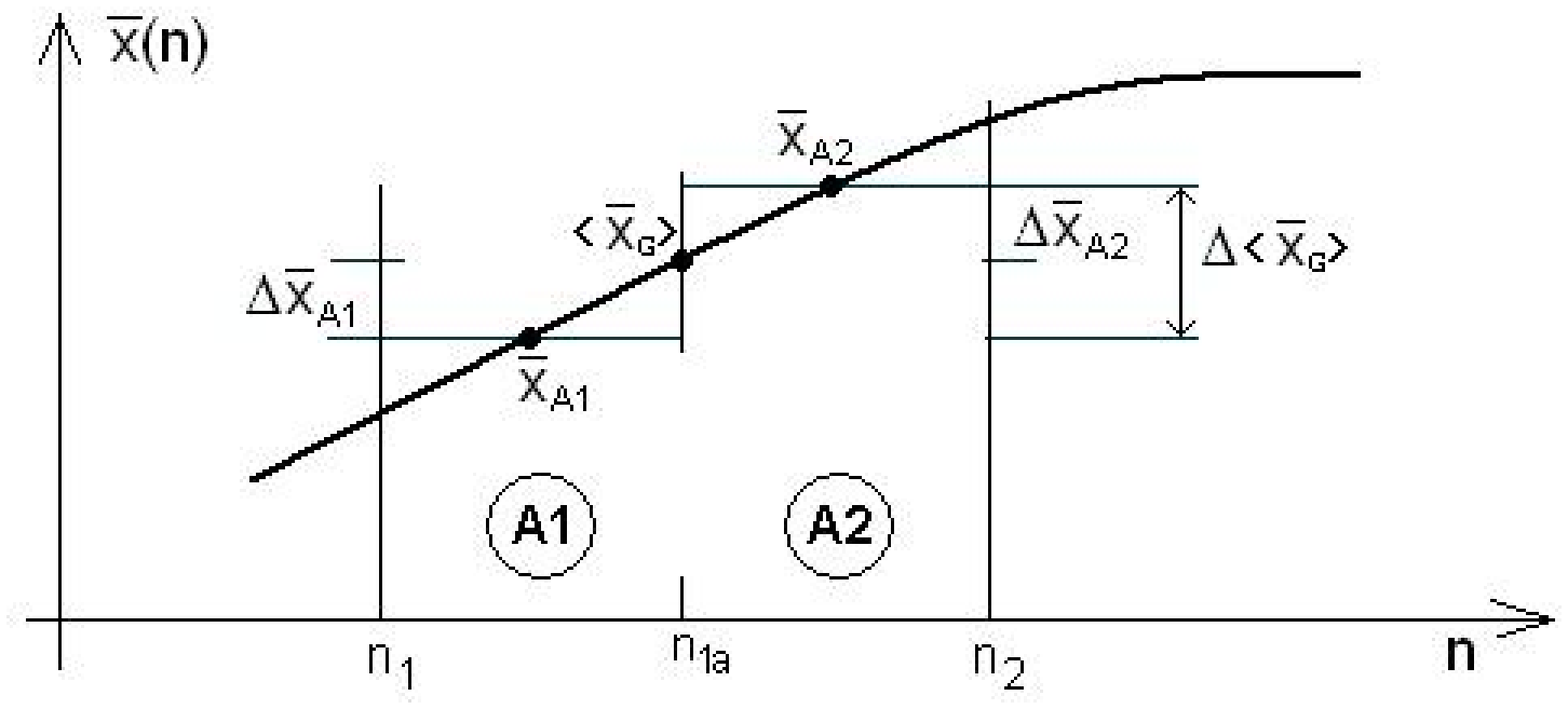}

{\small
{{\bf Fig.1:} Calculation of $C_2$ correlator using sub-samples $A1$ and $A2$.}  
}
\label{fig:Shapes1}
\end{figure}

Assuming $N_{\!A1}=N_{\!A2}=N_{ev}/2 $ and
$|\Delta \bar x_{\!A2}| = |\Delta \bar x_{\!A1}|$ (see Fig.1) one obtains
a simple relation
\begin{equation}
C_2^{A1+A2} = \frac{ C_2^{A1} + C_2^{A2}}{2}
              + 
              \frac{ (\Delta \langle \bar x_{G} \rangle)^2 }{4}
\end{equation}
where $\Delta \langle \bar x_{G} \rangle = 
\bar x_{A2} - \bar x_{A1} =
2 \Delta \bar x_{\!A2}$. 
If dynamics of particle production in sub-samples
$A1$ and $A2$ is identical then $C_2^{A1}\approx C_2^{A2}$ and one can express 
the exact "true" correlator 
$C_2^{\text{\,true}} = (C_2^{A1} + C_2^{A2})/2$ for the event sample $A$ as

\begin{equation}
C_2^{\text{\,true}} = C_2^{\,calc} - \frac{(\Delta \langle \bar x_{G} \rangle)^2}{4}
\end{equation}   

Correction term $(\Delta \langle \bar x_{G} \rangle)^2 / 4$ can be calculated and
subtracted from the correlator $C_2^{\,calc}$ which is
evaluated using global observable mean $\langle \bar x_G \rangle$ 
of the whole sample $A=A1+A2$.
Increasing the number of sub-samples $A1,A2,\ldots ,A_N$
gives (in general) a more precise correlator estimate. 
Using 4 subsamples of the event sample $A=A1+A2+A3+A4$\,
one obtains 
\begin{equation}
C_2^{A} = \frac{N_{\!A1}\cdot C_2^{A1} + N_{\!A2}\cdot C_2^{A2} 
              + N_{\!A3}\cdot C_2^{A3} + N_{\!A4}\cdot C_2^{A4} }{N_{\!A1} + N_{\!A2} + N_{\!A3} + N_{\!A4}}
              + 
              \frac{ (\Delta \langle \bar x_{G} \rangle)^2 }{4} [ 1+1/4 ]
\label{C2xx}
\end{equation}
if multiplicity dependence of observable mean $\bar x(n_k)$ 
increases in 4 steps: 
$\bar x(n_k\!\in\! A_1) = \bar x_{\!A1}$,\,
$\bar x(n_k\!\in\! A_2) = \bar x_{\!A2}$,\,
$\bar x(n_k\!\in\! A_3) = \bar x_{\!A3}$,\,
$\bar x(n_k\!\in\! A_4) = \bar x_{\!A4}$
(see Fig.2). 

\begin{figure}[h]
\includegraphics[width=10cm]{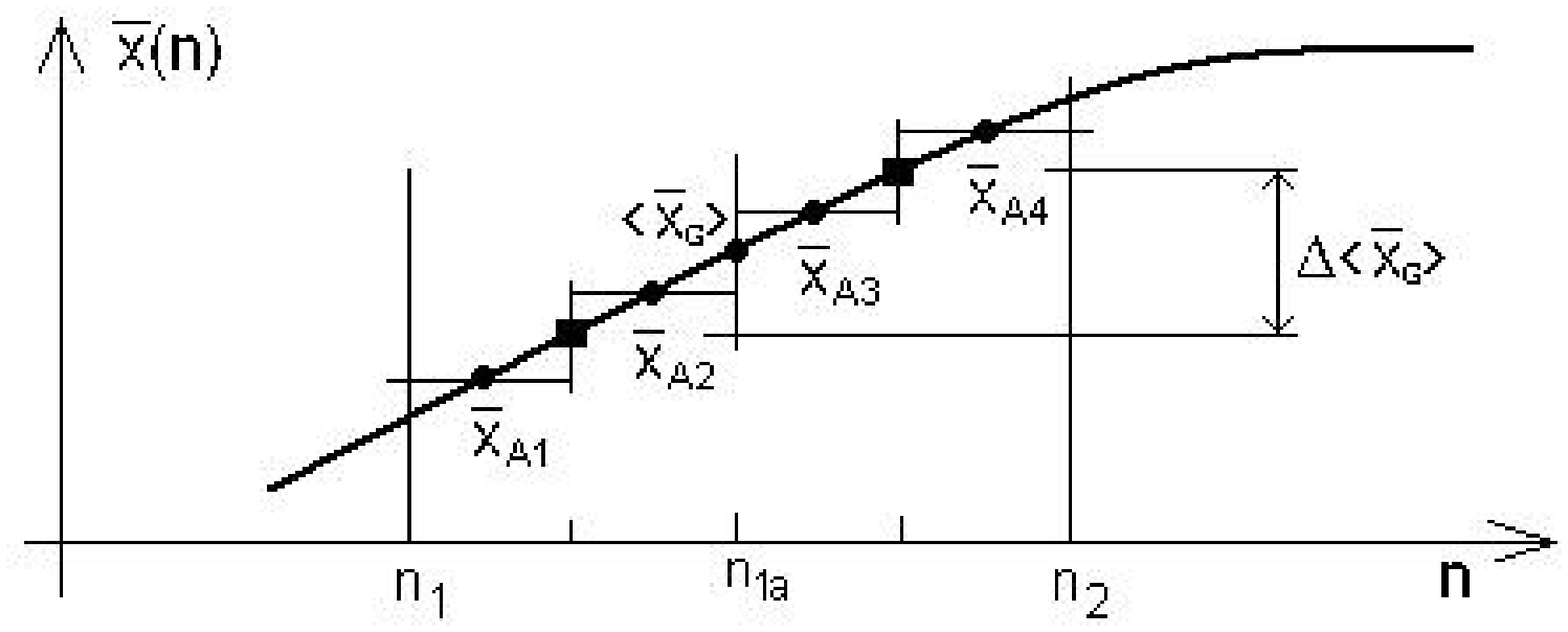}

{\small
{{\bf Fig.2:} Calculation of $C_2$ correlator using subsamples $A1+A2+A3+A4$.}  
}
\label{fig:Shapes2}
\end{figure}

The goal of the next section is to show how to evaluate correction terms 
for any given dependence of observable mean $\bar x_G(n_k)$ on observed 
event multiplicity $n_k$.

\section{Corrections to $C_2$ in General case}
Let us consider a general multiplicity-dependence of observable mean 
$\bar x(n_k)$.
After substitution 
$\langle \bar x_G \rangle \rightarrow \langle \bar x_G \rangle - \bar x(n_k) + \bar x(n_k)$
in Eq.(1) correlator $C_2^A$ is
\begin{equation}
C_2^A = C_2^{\text{\,calc}}\! = \frac{1}{N_{ev}}\sum_{N_{ev}}
\sum\!\!\!\!\!\!\!\sum_{\!\!i \neq j}^{\!\![N_{ij}]}
\frac{[x_i - \bar x(n_k) + \Delta \bar x_G(n_k)]
[x_j - \bar x(n_k) + \Delta \bar x_G(n_k)]}{N_{ij}}
\end{equation}
where $\Delta \bar x_G(n_k) = \bar x(n_k) - \langle \bar x_G \rangle$.
Here observable mean $\bar x(n_k)$ of events with measured multiplicity $n_k$ is defined as
\begin{equation} 
\bar x(n_k) = \sum_{N=1}^{N_{n_k}} \bigg[ \sum_{i=1}^n \frac{x_i}{n} \bigg] / N_{n_k}
\label{OBm}
\end{equation}
where $N_{n_k}$ is total number of events in sample A with measured multiplicity $n_k$
and $n$ denotes number of particles (in each event) used for the calculation of correlators. 
Since $\sum^{N_{n_k}}_{N=1}\!\sum_{i=1}^{n}(x_i - \bar x(n_k)) = 0$  
the expression becomes
\begin{equation}
C_2^{\,\text{calc}} = \frac{1}{N_{ev}}\sum_{N_{ev}}
\sum\!\!\!\!\!\!\!\sum_{\!\!i \neq j}^{\!\![N_{ij}]}
\frac{(x_i - \bar x(n_k))(x_j - \bar x(n_k))}{N_{ij}}
      + \sum_{N_{ev}}\frac{(\Delta \bar x_G(n_k))^2}{N_{ev}}
\label{crrX}
\end{equation}

The first term on the right-hand side of Eq.(\ref{crrX}) is a correlator
calculated using a multiplicity-adjusted 
observable mean $\bar x(n_k)$. We call this "true" correlator 
$C^{\text{\,true}}_2$  which coincides formally with global
correlator defined by Eq.(\ref{C2x}) evaluated for events with fixed
multiplicity $n_k$. Such correlator does not contain correction terms of type given 
by Eq.(7) due to zero width ($n_2=n_1=n_k$) of multiplicity interval of events being analyzed.  
True correlator
\begin{equation}
C_2^{\text{\,true}} = \frac{1}{N_{ev}}\sum_{N_{ev}}
\sum\!\!\!\!\!\!\!\sum_{\!\!i \neq j}^{\!\![N_{ij}]}
\frac{(x_i - \bar x(n_k))(x_j - \bar x(n_k))}{N_{ij}}
\end{equation}
is also a limiting case of the first term on the right-hand side of
Eq.(\ref{C2xx}):
\begin{equation}
C_2^{\text{\,true}} = 
\lim _{k->\Delta n} \frac{\sum_{i=1}^k N_i\cdot C^{Ai}_2}{\sum_{i=1}^k N_i} 
\end{equation}
where $\Delta n = n_2-n_1$ is the maximal number of multiplicity 
sub-intervals in the sample $A$.
Using a more explicit notation one has
\begin{equation}
C_2^{\text{\,true}}(\bar x(n_k)) = C_2^{\text{\,calc}}(\langle \bar x_G \rangle)
     - \sum_{N_{ev}} \frac{[\bar x(n_k) - \langle \bar x_G \rangle]^2}{N_{ev}}
\label{C2x15}
\end{equation}
where correction term $\sum (\bar x(n_k) - \langle \bar x_G \rangle)^2/N_{ev}$ 
can be further expressed as
\begin{equation}
 \sum_{N_{ev}} \frac{[\bar x(n_k) - \langle \bar x_G \rangle]^2}{N_{ev}}
=\sum_{n_k=n_1}^{n_2} \frac{N_{n_k} 
                     [\bar x(n_k) - \langle \bar x_G \rangle]^2}{N_{ev}}
=\int_{n_1}^{n_2} \! P(n_k)\cdot [\bar x(n_k) - \langle \bar x_G \rangle]^2 dn_k
\label{C2x16}
\end{equation}
Here $N_{n_k}$ is the number of events with multiplicity $n_k$ in sample $A$ and 
probability density $P(n_k)$ for events with multiplicity $n_k$ is
$P(n_k) = N_{n_k}/\sum N_{n_k}$. Based on
Eq.(\ref{C2x15}) one can find "true" correlator 
for the event sample $A$ by subtracting
the correction term defined by Eq.(\ref{C2x16})  
from global correlator $C_2^{\text{\,calc}} = C_2^A(\langle \bar x_G \rangle)$ 
(calculated using the global observable mean
$\langle \bar x_G \rangle = \sum \bar x_G(n_k) N_{n_k}/N_{ev}$).
To summarize our result in analytical form we express "true"
2-particle correlator as
\begin{equation} 
C_2^{\text{\,true}} (\bar x(n_k)) = 
                      C_2^{\text{\,calc}}(\langle \bar x_G \rangle) 
- \int_{n_1}^{n_2} \! P(n_k)\cdot [\bar x(n_k) - \langle \bar x_G \rangle]^2 dn_k
\label{C2int}
\end{equation}
where meaning of $P(n_k)$, $\langle \bar x_G \rangle$ and $\bar x(n_k)$ has been
described in the text above.
Assuming linear multiplicity dependence of observable mean
($\bar x(n_k) = x_0 + \tilde k\cdot n_k$) and constant $P(n_k)$ distribution
one evaluates integral in Eq.(\ref{C2int}) as\,
$\tilde k^2(n_2-n_1)^2/12
 = (\Delta \langle \bar x_G\rangle)^2/3$.
This is in agreement with Eq.(\ref{C2xx}) suggesting corrections 
$[1+1/4+1/16+\!\cdots\,]
(\Delta\langle\bar x_G\rangle)^2/4\,$ if number of subsamples $\{A_N\}$ is
iteratively doubled.
(Note that $\sum_{n=0}^\infty 1/2^{2n}$ = 4/3.)

\section{Corrections for $C_K$ Correlators}
Results obtained for $C_2$ correlator can be 
generalized for higher order correlators.
Let us define $K$-th order global correlator $C_K$ 
\begin{equation}
C_K^{(n,x)}(\langle \bar x_G \rangle) =
\frac{1}{N_{ev}}\sum_{N_{ev}} \!\!\!\!\!\!\!\!\!\!\!
\sum_{\ \ \ \ \ \ \ i_1\neq i_2 \cdots \neq i_K}^{
      \ \ \ \ \ \ \ [N_{i_1 i_2\cdots i_K}]}
                \!\!\!\!\!\!\!\!\!\!\!\!\!\!\!\!
                \cdots \!
                \sum \,\,\,
\frac{\prod_{m=1}^K(x_{i_m}\!\! - \langle \bar x_G \rangle)}{N_{i_1 i_2\cdots i_K}}
\label{globC}
\end{equation}
where $N_{i_1 i_2\cdots i_K}\!\! = n!/(n-K)!$\, is the number of 
particle $K$-plets $\{i_1,i_2\ldots i_K\}$
made from $n<n_k$ particles in each event ($n_k$ is total multiplicity of a given event).
After substitution 
$\langle \bar x_G \rangle\! =\! \bar x(n_k) - \Delta \bar x_G(n_k)$ (where
$\Delta \bar x_G(n_k)\! =\! \bar x(n_k)-\langle \bar x_G\rangle$) 
a compact expression can be
found 

\begin{equation}
C_K^{\,\text{calc}}(\langle \bar x_G\rangle) 
  = \sum_{\lambda = 0}^K \sum_{N_{ev}}  
    \frac{(\Delta \bar x_G(n_k))^{K-\lambda} \,\,
     C_\lambda^{\,\text{true}}(\bar x_G(n_k))}{N_{ev}}
    { K \choose \lambda }
\end{equation}
where $C_\lambda^{\,\text{true}}(\bar x(n_k))$ are $\lambda$-th order 
"true" correlators (defining $C_0^{\,\text{true}}(\bar x(n_k)) = 1$)
formally obtained by replacing $\langle \bar x_G \rangle$ in Eq.(\ref{globC})
by $\bar x(n_k)$.
In particular, for $N=3$ one has
\begin{equation}
C_3^{\text{\,calc}} (\langle \bar x_G \rangle) = 
          C_3^{\text{\,true}}( \bar x(n_k))
+3\sum_{N_{ev}}\frac{\Delta\bar x(n_k) \cdot C_2^{\text{\,true}}(\bar x(n_k))}{N_{ev}}
                    + \sum_{N_{ev}}\frac{(\Delta \bar x(n_k))^3}{N_{ev}}
\label{C3sm}
\end{equation} 
since $C_1^{\,\text{true}}(\bar x(n_k))\!=\!0\,$ 
due to definition of $\bar x(n_k)$. For constant 
probability $P(n_k)$ 
and for function $\Delta \bar x_G(n_k)$ antisymmetric 
around average $\langle n_k\rangle$
term $(\Delta \bar x_G(n_k))^{3}$
becomes zero. Moreover, if  
$C_2^{\text{\,true}}(\bar x(n_k))$ is a constant function of $n_k$ one obtains
$C_3^{\text{\,calc}} (\langle \bar x_G \rangle) = C_3^{\text{\,true}}( \bar x(n_k))$.
Under the same assumptions for $N=4$ one has
\begin{equation}
C_4^{\,\text{calc}}(\langle \bar x_G \rangle) =
C_4^{\,\text{true}}(\bar x(n_k)) \, + \, 
\sum_{N_{ev}}\bigg[
\frac{(\Delta \bar x_G(n_k))^4}{N_{ev}}
+6\frac{(\Delta \bar x_G(n_k))^2\, C_2^{\,\text{true}}(\bar x(n_k))}{N_{ev}}
\bigg]
\label{C4sm}
\end{equation}
and for $N=5$
\begin{equation}
C_5^{\,\text{calc}}(\bar x(n_k)) = 
C_5^{\,\text{true}}(\langle \bar x_G\rangle) 
   + 10 \sum _{N_{ev}}\frac{(\Delta \bar x_G(n_k))^2\, 
     C_3^{\,\text{true}}(\bar x(n_k))}{N_{ev}}
\label{C5sm}
\end{equation}
where $C_\lambda^{\,\text{true}}(\bar x(n_k))$ are the $\lambda$-th order
true correlators defined as
\begin{equation}
C_\lambda^{\text{true}}(\bar x(\tilde n_k)) =
\frac{1}{N_{ev}}\sum_{N_{ev}} \!\!\!\!\!\!\!\!\!\!\!
\sum_{\ \ \ \ \ \ \ i_1\neq i_2 \cdots \neq i_\lambda}^{
      \ \ \ \ \ \ \ [N_{i_1 i_2\cdots i_\lambda}]}
                \!\!\!\!\!\!\!\!\!\!\!\!\!\!\!\!
                \cdots \!
                \sum \,\,\,
\frac{\prod_{m=1}^\lambda
(x_{i_m}\!\! - \bar x(\tilde n_k))}{N_{i_1 i_2\cdots i_\lambda}}
\label{CtrueDef}
\end{equation}

We have thus obtained expressions for
differences between 
$C_K^{\text{\,calc}}=C_K^{\text{\,calc}}(\langle x_G\rangle)$
correlators calculated using global 
observable mean $\langle \bar x_G \rangle$
and true correlators $C_K^{\text{\,true}}=C_K^{\text{\,true}}(\bar x(n_k))$
evaluated using multiplicity-adjusted observable mean $\bar x(n_k)$:
\begin{equation}
C_K^{\text{\,calc}} = C_K^{\text{\,true}} + \,\,\Delta_K
\end{equation}
Correction term $\Delta_2$ according to Eq.(\ref{C2int}) is
\begin{equation}
     \Delta_2 = \sum_{N_{ev}} \frac{(\Delta \bar x_G(n_k))^2}{N_{ev}}
 =\int_{n_1}^{n_2} \! P(n)\, [\,\bar x(n) - \langle \bar x_G \rangle\,]^2 dn
\label{D2x}
\end{equation}
and from Eq.(\ref{C4sm}) one has
\begin{equation} \Delta_4 =
  \int_{n_1}^{n_2} \! P(n)\, [\,\bar x(n) - \langle \bar x_G \rangle\,]^4 dn
+6\int_{n_1}^{n_2} \! P(n)\, C_2^{\,\text{true}}(\bar x(n))\,
                             [\,\bar x(n) - \langle \bar x_G \rangle\,]^2 dn
\label{D4x}
\end{equation}
For 
 $P(n_k)$ symmetrical around $\langle n_k \rangle$ term
$\Delta_3 \rightarrow 0$. However, correction $\Delta_5$ 
remains non-zero
\begin{equation}
\Delta_5  = 10
\int _{n_1}^{n_2} \! P(n)\, C_3^{\,\text{true}}(\bar x(n))\,
[\,\bar x(n) - \langle \bar x_G \rangle\,]^2\, 
dn
\label{D5x}
\end{equation}
For probability distribution ${\cal P}(n_k)$ {\it {asymmetrical\,}} around
$\langle n_k \rangle $ one obtains from Eq.(\ref{C3sm})
\begin{equation}
\Delta_3 =
  \int_{n_1}^{n_2} \! {\cal P}(n)\, [\,\bar x(n) - \langle \bar x_G \rangle\,]^3 dn
 +\int_{n_1}^{n_2} \! {\cal P}(n)\, C_2^{\text{\,true}}(\bar x(n))\,
                      [\,\bar x(n) - \langle \bar x_G \rangle\,] dn
\label{D3x}
\end{equation}

In a real calculation one does not have to calculate explicitly the
multiplicity-dependent observable mean $\bar x(n_k)$ 
for each multiplicity $n_k$.
It is enough to split event sample $A$ into reasonable number of subsamples
$A_1,A_2,\cdots A_N$ and calculate mean $\bar x[A_N]$ for subsamples $A_N$.
Fitting $\bar x[A_N]$ values with smooth function $\bar X(n_k)$ gives
approximation for $\bar x(n_k)$.

One can evaluate
"true" correlators also 
directly using multiplicity-adjusted observable mean $\bar x(n_k)$
for subsets of events with fixed multiplicities $n_k$.
This method has been utilized in $p_t$
correlations analysis \cite{STAR}. However, there is
still another systematical effect which should be accounted for
if one wants to obtain correlator values free from contributions
due to multiplicity-dependent observable mean $\bar x(n_k)$.
We discuss this issue in the next section.

\section{Fixed-multiplicity corrections to correlators}
Let us consider now the situation when multiplicity 
of particles $n_k$ in a given event 
is not known precisely and instead of $n^{\text{tot}}$ 
(total number of particles produced in a given event)
a multiplicity of tracks $\tilde n_k \approx n^{\text{tot}}/\xi\,$ 
is measured in a detector. In events with a given fixed 
measured multiplicity $\tilde n_k$
there will be fluctuations of corresponding $n_k^{\text{tot}}$ 
around the average value
$\langle n^{\text{tot}} \rangle \approx \tilde n_k \cdot \xi\,$ 
(where $\xi > 1$ is a real number
corresponding to the detector acceptance). 
If observable mean
of quantity under study does not depend on multiplicity
($\bar x(n) = \text{\it{const}}\,$)
there is no influence on correlator values from these fluctuations.

However, if
observable mean $\bar x(n)$
depends on multiplicity additional corrections $\tilde \Delta_n$ 
to calculated correlator values appear due to $n_i^{\text{tot}}$ 
fluctuations at fixed measured $\tilde n_k$ 
(which generate fluctuations of observable mean 
$\bar x(n_i^{\text{tot}})$ values at given $\tilde n_k$). 

Contributions $\tilde \Delta_n$ are the 
{\it{fixed-multiplicity}} corrections to correlators.
They can influence results and interpretations of the correlator analysis
if they are not accounted for.

For large enough measured multiplicities $\tilde n_k$ 
fluctuations of measured multiplicity $\tilde n_k$ at given fixed 
total multiplicity $n^{\text{tot}}$ are close to Gaussian (see Appendix) with
probability distribution $P(\tilde n_k | n^{\text{tot}})$:
\begin{equation}
P(\tilde n_k | n^{\text{tot}}) = 
\frac{e^{-(\tilde n_k - \langle \tilde n_k\rangle)^2/2\sigma^2_{\tilde n_{\!k}}} }
     {\sqrt{2\pi \sigma^2_{\tilde n_{k}}\,}}
\label{Gnk1}
\end{equation}
where $\langle \tilde n_k \rangle \approx n^{\text{tot}}/\xi\,$ and
$\,\sigma_{\tilde n_{k}} = c\cdot \sqrt{(n^{\text{tot}}/\xi)}$ (see Appendix).
One can express probability of $n^{\text{tot}}$
fluctuations $P(n^{\text{tot}}|\tilde n_k)$ at given fixed measured 
$\tilde n_k$ using Bayes' theorem \cite{Papoulis}.
For constant $P(n^{\text{tot}})=const\,$ probability distribution
one obtains (see Appendix)
\begin{equation}
P(n^{\text{tot}}|\tilde n_k) = 
\frac{
e^{-(n^{\text{tot}} - \tilde n_k\cdot \xi)^2/2\sigma^2_{\text{tot}} } }{ 
      \sqrt{2\pi \sigma^2_{\text{tot}}\, } }
\label{PnTot1}
\end{equation}
where $\sigma_{\text{tot}} = \xi\cdot \sigma_{\tilde n_k}$.
Assuming linear approximation 
$\bar x(n^{\text{tot}}) = x_0 + \tilde k\cdot n^{\text{tot}}$ for
multiplicity-dependent observable mean $\bar x(n^{\text{tot}})$
in $3\sigma_{\text{tot}}$ vicinity of  
$n^{\text{tot}} \approx \tilde n_k\cdot\xi\,$ one obtains 
fixed-multiplicity correction term $\tilde \Delta_2$ from Eq.(\ref{D2x}) as 
\begin{equation}
\tilde \Delta_2 = \int 
\! P(n^{\text{tot}}|\tilde n_k)\,
[\bar x(n^{\text{tot}})-\bar x(\langle n^{\text{tot}}\rangle_{\tilde n_k})]^2 \,
dn^{\text{tot}}\! = 
\tilde k^2 \int 
\frac{(n-\xi\!\cdot\!\tilde n_k)^2\, 
      e^{-(n-\xi\cdot \tilde n_k)^2/2\sigma_{\text{tot}}^2}}
     {\sqrt{2\pi \sigma_{\text{tot}}^2\,}}\, dn
\label{X25}
\end{equation}
This gives a simple result
\begin{equation}
\tilde\Delta_2=
\tilde k^2\cdot\sigma_{\text{tot}}^2(\xi)
\label{DT2a}
\end{equation}
where 
$\sigma_{\text{tot}}(\xi) = \sigma_{\tilde n_k}\xi
 =\xi \sqrt{\tilde n_k}\,\langle\tilde \sigma \rangle $ (see Appendix).
For slope parameter $k$ obtained from approximation
$\bar x(\tilde n_k) = x_0 + k\cdot \tilde n_k\,$ (measured
experimentally) one has 
$\tilde k = k/\xi \,\,$ and thus
$\tilde \Delta_2 = 
  k^2\, \tilde n_k\, \langle \tilde \sigma \rangle^2$
where $\langle \tilde \sigma \rangle\approx 1$ 
is to be obtained from MC simulation.
Using similar arguments 
correction term $\tilde \Delta_4$ can be calculated 
from Eq.(\ref{D4x}) as
\begin{equation}
\tilde\Delta_4 = 3\,\tilde k^4\!\cdot \sigma_{\text{tot}}^4(\xi) 
      +6\,\tilde k^2\!\cdot\sigma_{\text{tot}}^2(\xi)\,C_2^{\,\text{true}}
\label{DT4a}
\end{equation}
Acceptance parameter $\xi$ disappears ($\tilde k\!=\!k/\xi$ and 
$\sigma_{\text{tot}}= \sigma_{\tilde n_k}\xi$):
$\tilde\Delta_4 = 3\,k^4\sigma^4_{\tilde n_k} 
                + 6\,k^2\sigma^2_{\tilde n_k}C_2^{\text{\,true}}$
(where $\sigma_{\tilde n_k}
 =\sqrt{\tilde n_k}\,\langle\tilde \sigma \rangle $ see Appendix).
Fixed-multiplicity correction terms thus
depend only on experimentally measurable quantities.

One might be tempted to imply
$\tilde \Delta_3 \longrightarrow 0$
based on symmetrical probability distribution given by Eq.(\ref{PnTot1}). 
However,
fluctuations of ($n^{\text{tot}}_i - \langle n^{\text{tot}}\rangle$) values
can be significantly asymmetrical for $\tilde n_k$ small enough 
(see Fig.3 for $n^{\text{tot}}_i$ fluctuations at $\tilde n_k=10$). 

\begin{figure}[h]
\includegraphics[width=7.4cm]{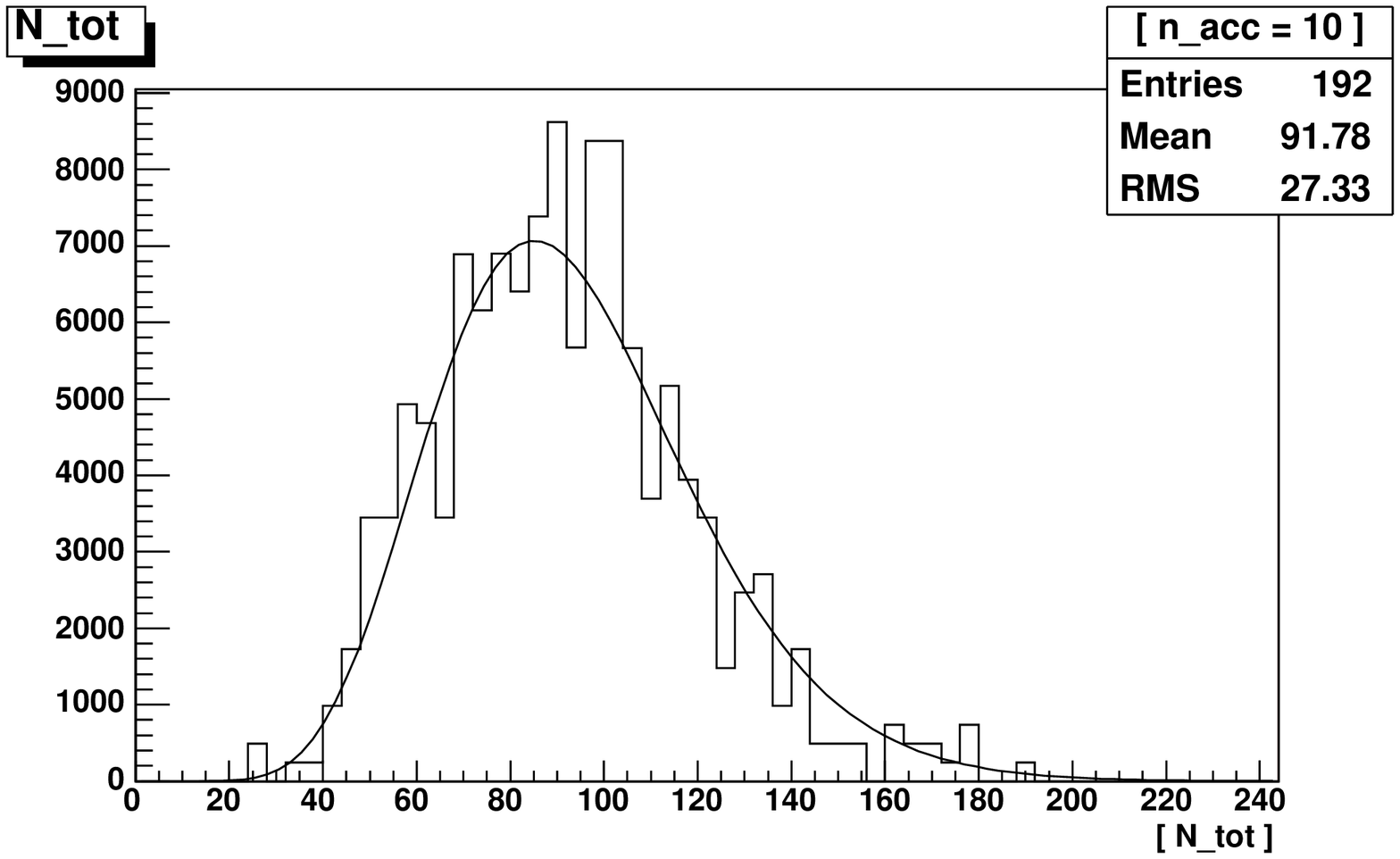}
\includegraphics[width=7.4cm]{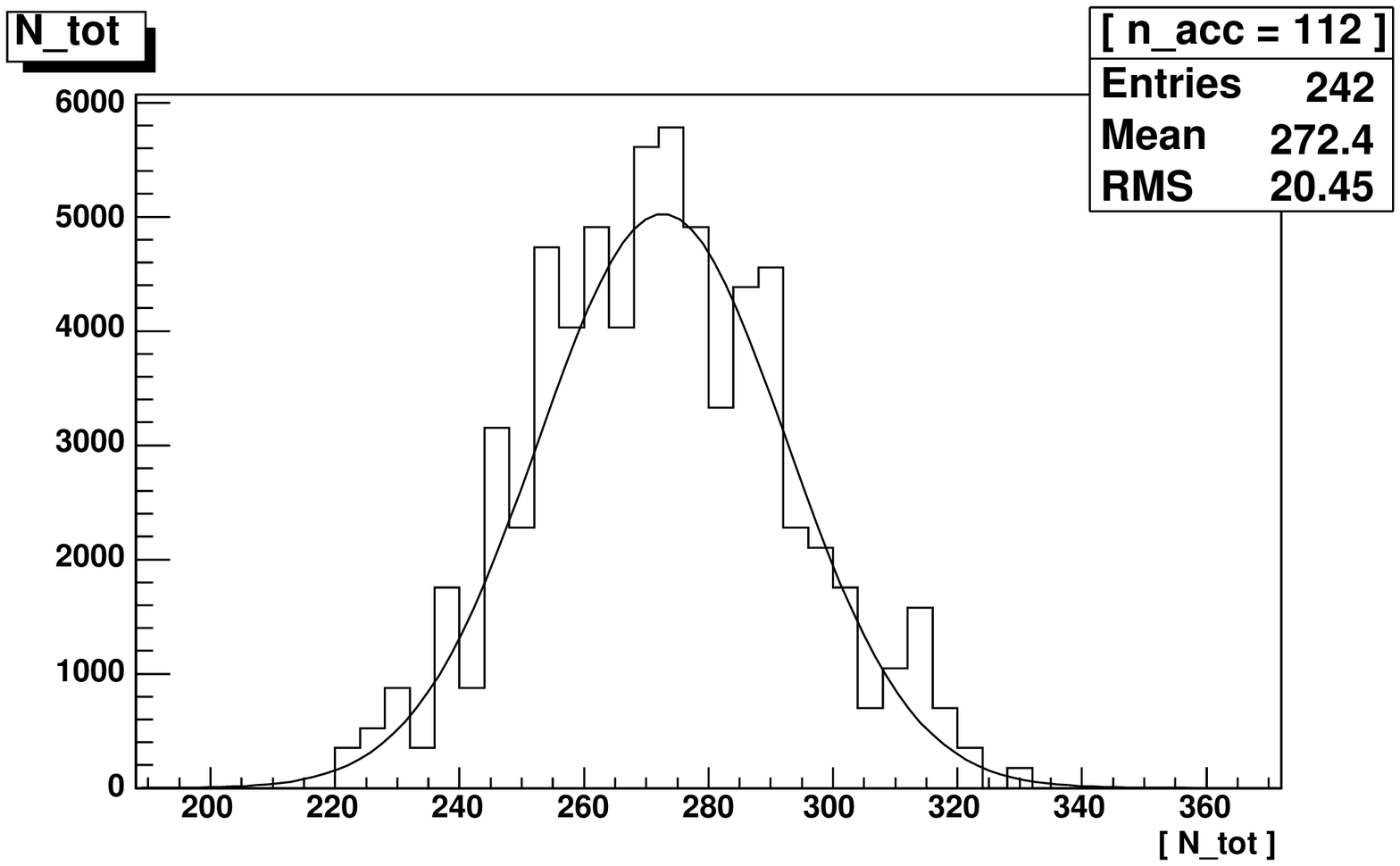}

{\small 
{{\bf Fig.3:} Fluctuations of $n^{\text{tot}}_i$ for $\,\tilde n_k$=10, $\xi$=8.51
and  $\tilde n_k$=112, $\xi$=2.44 (MC simulation).}
}
\label{fig:Shapes3}
\end{figure}

In this case 
correction $\tilde \Delta_3$
can be evaluated using Eq.(\ref{D3x}).
Symmetrical $P(n^{\text{tot}}|\tilde n_k)$  given by Eq.(\ref{Gnk1}) 
yields $\tilde \Delta_3 = 0$ and from
Eq.(\ref{D5x}) one obtains
$\tilde \Delta_5 = 10\,k^2\sigma_{\tilde n_k}^2C_3^{\,\text{true}}\,$.

\section{MC Simulations}

In order to verify behavior of $\sigma_{\text{tot}}(\xi)$
and $\sigma_{\tilde n_k}(\tilde n_k)$ 
a simple MC simulation has been performed: 
Events with total multiplicities $n^{\text{tot}}_k\in (50,2000)$
have been generated with constant probability 
$P(n^{\text{tot}})\! =\! \text{\it{const}}$.
In each event, rapidities $y_i$ were assigned to $n^{\text{tot}}_k$ particles according to
the bell-shaped rapidity distribution and number of observed
particles $\tilde n_k$ 
found in the selected acceptance range $(-y_a,y_a)$ has been determined.

Two-dimensional histogram H2$(\tilde n_k,n^{\text{tot}}_k)$ filled with pairs
of obtained numbers $n^{\text{tot}}_k$ 
and $\tilde n_k$ is shown in Fig.4. 
Projection histograms H$_{n^{\text{tot}}_k}(\tilde n_k)$
and H$_{\tilde n_k}(n^{\text{tot}}_k)$ which
are proportional to probabilities $P(\tilde n_k|n^{\text{tot}})$ from Eq.(\ref{AXpnk})
and $P(n^{\text{tot}}|\tilde n_k)$
given by Eq.(\ref{AXpntot}) 
are shown for $\xi \approx 4.26$.

\begin{figure}[h]
\includegraphics[width=5.0cm]{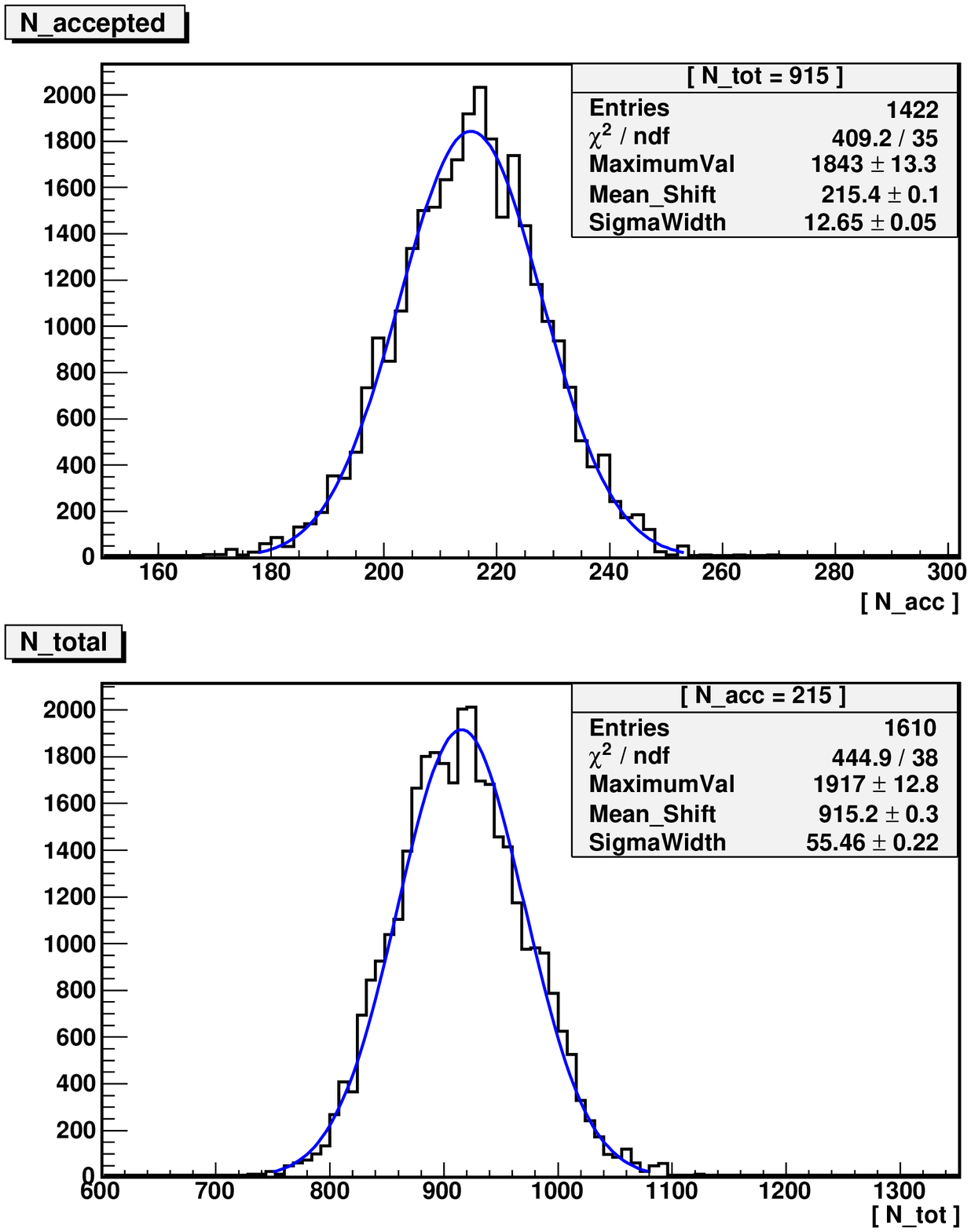}
\includegraphics[width=4.9cm]{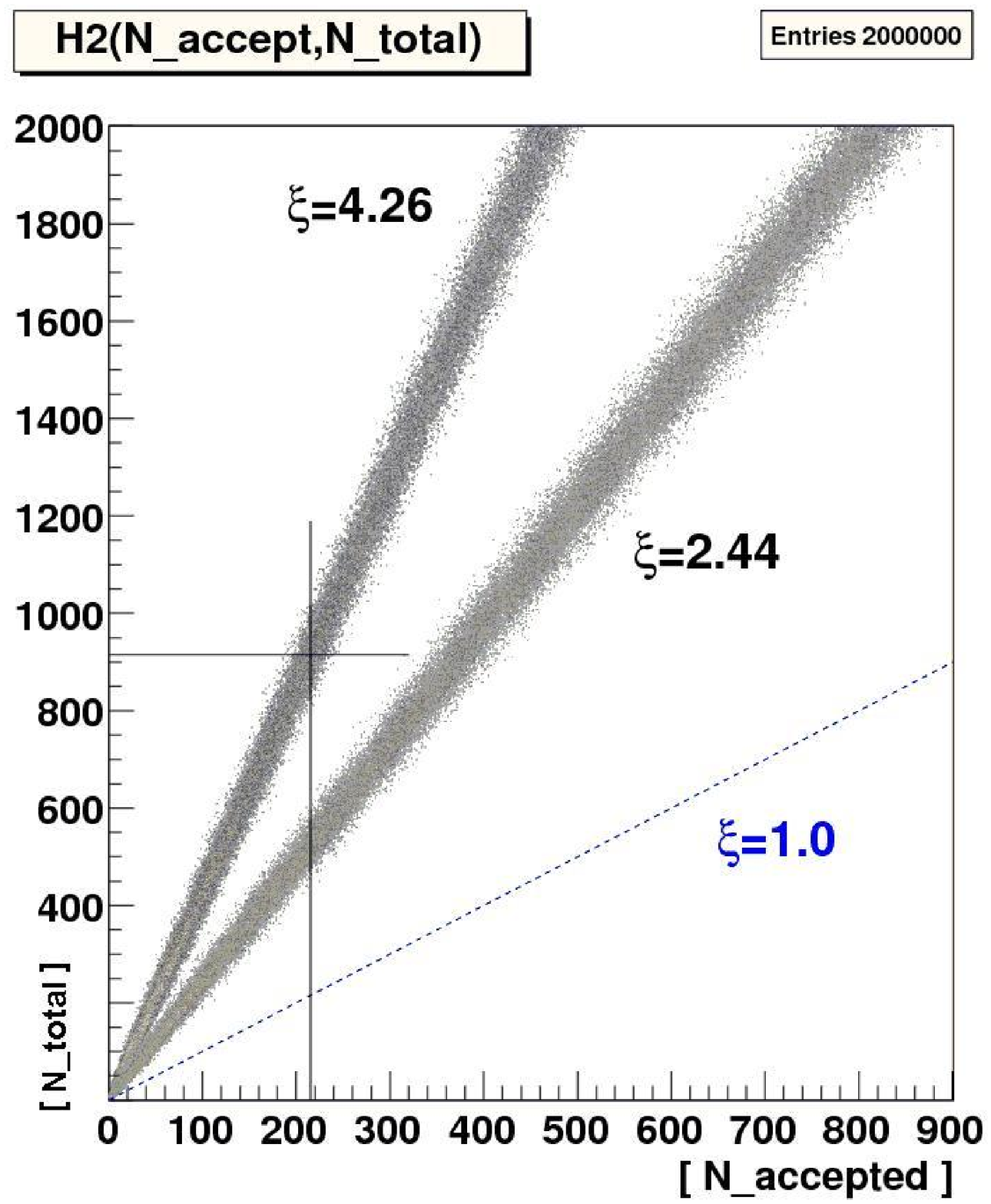}
\includegraphics[width=5.0cm]{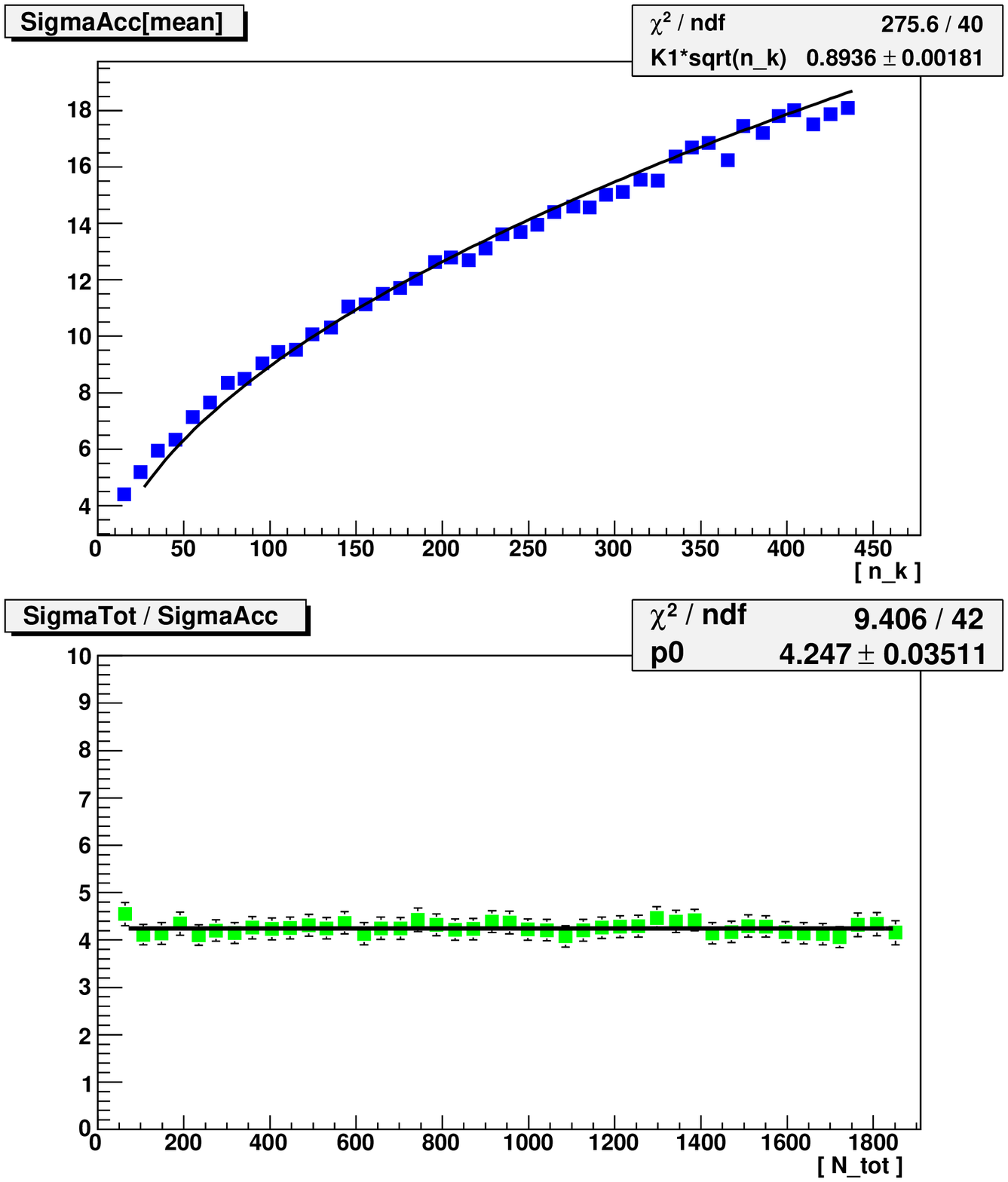}

{\small
{{\bf Fig.4:} Histograms H$_{\tilde n_k}(n^{\text{tot}}_k)$,
                         H$_{n^{\text{tot}}_k}(\tilde n_k)$,
                         H2$(\tilde n_k,n^{\text{tot}}_k)$,
            fitted $\sigma_{\tilde n_k}\!= c\cdot \sqrt{n_k}$
         and ratio $\sigma_{\text{tot}}/\sigma_{\tilde n_k}$.
            }
}
\label{fig:Shapes4}
\end{figure}
Widths $\sigma_{\text{tot}}(\xi,\tilde n_k^*)$ 
and $\sigma_{\tilde n_k}(n^{\text{tot}}_*\!=\!\xi \cdot n_k^*)$ 
have been obtained from Gaussian fits of the projection histograms.
Width $\sigma_{\tilde n_k}$ has been found to follow
$c\,\sqrt{\tilde n_k}$ dependence
with $c\approx 0.9$ in agreement with Eq.(\ref{X51}).
Ratio of widths $\sigma_{\text{tot}}(\xi,\tilde n^*_k)$ 
and $\sigma_{\tilde n_k}(n^{\text{tot}}_*)$ 
at $n^{\text{tot}}_*\! \approx \xi \cdot \tilde n_k^*\,$
appeared to be constant and equal to 
$\,\sigma_{\text{tot}}(\xi,n^{\text{tot}}_*/\xi)/
\sigma_{\tilde n_k}(n^{\text{tot}}_*)\approx \xi\,$
as expected from Eq.(\ref{AXpntot}).

Simulations for $P(n^{\text{tot}})\neq const\,$ have also been done and shift
$\delta n^{\text{tot}}\!=\langle n^{\text{tot}}\rangle_{\tilde n_k} 
-\tilde n_k\!\cdot\!\xi\,$ 
(described in the next secion) has been observed.

\section{Fixed-multiplicity Corrections in real experiments} 
In a general case probability
$P(n^{\text{tot}}) \neq \text{const}$ 
(usually $P(n^{\text{tot}})$ decreases with $n^{\text{tot}}$).
The situation is slightly more complex in this case
and there appears a shift of average $\langle n^{\text{tot}}\rangle_{\tilde n_k}$
(evaluated for events with fixed measured $\tilde n_k$)
relative to $\langle \tilde n^{\text{tot}} \rangle = \tilde n_k\cdot \xi$
expected in the case of $P(n^{\text{tot}})=\text{const}$.
We will show that the shift
\begin{equation}
\delta n^{\text{tot}} = 
\langle n^{\text{tot}}\rangle_{\tilde n_k} - \tilde n_k\cdot \xi
\end{equation}
does not influence fixed-multiplicity correction terms for correlators significantly.
Let us assume $P(n^{\text{tot}})$
decreasing with $n^{\text{tot}}$ as
$a_{\text{o}}\! - a_1 \cdot n^{\text{tot}}$ which gives
(denoting $a_1/a_{\text{o}} = \bar\omega $) normalized prior probability
$P(n^{\text{tot}})$ in the form\footnote{This probability distribution
implies $n^{\text{tot}}<1/\bar\omega$.}
\begin{equation}
P(n^{\text{tot}})=
 2\bar\omega(1-\bar\omega \cdot n^{\text{tot}})
\label{PomegaB}
\end{equation}

For a given fixed $n^{\text{tot}}$ the average
$\langle \tilde n_k\rangle = n^{\text{tot}}/\xi\,$ and
if $\langle \tilde n_k\rangle$ is large enough
fluctuations of $\tilde n_k$
are close to Gaussian according to Eq.(\ref{AXpntot}).
However, average value $\langle n^{\text{tot}} \rangle$
for a given fixed measured $\tilde n_k$ is
\begin{equation}
\langle n^{\text{tot}}\rangle_{\tilde n_k} = 
   \int
   \! n^{\text{tot}} P(n^{\text{tot}} | \tilde n_k) d n^{\text{tot}}
\label{AVntot}
\end{equation}
where $P(n^{\text{tot}} | \tilde n_k )$ is a conditional probability distribution 
of having event with total multiplicity $n^{\text{tot}}$ (for events
with fixed measured multiplicity $\tilde n_k$). Using prior probability 
$P(n^{\text{tot}})$ given 
by Eq.(\ref{PomegaB}) $P(n^{\text{tot}}|\tilde n_k)$ can be found analytically
if $\,\xi\,\tilde n_k+3\,\sigma_{\tilde n_k}\xi < 1/\bar\omega$ (see Appendix)

\begin{equation}
P(n^{\text{tot}}|\tilde n_k)
=\frac{e^{-(n^{\text{tot}} -\tilde n_k\cdot \xi)^2/2\xi^2\sigma^2_{\tilde n_k}} }{
       \sigma_{\tilde n_k}\sqrt{2\pi\,}\,\xi } 
 \frac{(1-n^{\text{tot}} \bar\omega )}{
       (1-\tilde n_k\xi\, \bar\omega )}
\label{PntotOmega}
\end{equation}
and Eq.(\ref{AVntot}) yields
\begin{equation}
\langle n^{\text{tot}}\rangle_{\tilde n_k} = \xi\cdot\tilde n_k -
\frac{\bar\omega\,\xi^2\sigma^2_{\tilde n_k}}{1-\tilde n_k\xi\,\bar\omega}
\label{NtotNk}
\end{equation}

One can also define 
{\it {effective}} observed multiplicity 
$\langle\tilde n_k\rangle_{\tilde n_k} 
= \langle n^{\text{tot}}\rangle_{\tilde n_k}/\xi\,$ for 
events with fixed measured $\tilde n_k$
\begin{equation}
\langle \tilde n_k \rangle_{\tilde n_k} =\,
\tilde n_k 
- \frac{\bar\omega\,\xi\,\sigma^2_{\tilde n_k}}{1-\tilde n_k\xi\,\bar\omega} =\,
\tilde n_k
- \frac{\omega\sigma^2_{\tilde n_k}}{1-\tilde n_k\omega}
\label{NkNk}
\end{equation}

Using slope parameter $\omega $ determined from 
$P(\tilde n_k) = 2\,\omega(1-\omega \cdot \tilde n_k)$ distribution
(accessible experimentally)
one has $\bar\omega = \omega/\xi$ comparing to Eq.(\ref{PomegaB}) and
parameter $\xi $ disappears from Eq.(\ref{NkNk}). 
Displacement 
$\delta \tilde n_k = \langle \tilde n_k \rangle_{\tilde n_k}\! -\, \tilde n_k\,$
of the effective multiplicity $\langle \tilde n_k \rangle_{\tilde n_k}$
relative to fixed measured multiplicity $\tilde n_k$
can be relevant e.g. in elliptic flow analysis where effective
participant eccentricity $\varepsilon_{\langle \tilde n_k \rangle}$
is compared to elliptic flow strength $v_2(\tilde n_k)$.

Fixed-multiplicity 
correction term $\tilde \Delta_2^\omega$ 
for prior probability
$P(n^{\text{tot}})$ given by Eq.(\ref{PomegaB}) can be evaluated 
analogously to Eq.(\ref{X25})
\begin{equation}
\tilde \Delta_2^{\omega} = \frac{k^2}{\xi^2}
\int P(n^{\text{tot}}|\tilde n_k) 
[n^{\text{tot}} - \langle n^{\text{tot}}\rangle_{\tilde n_k}]^2 dn^{\text{tot}} 
\end{equation}
which for shifted $\langle n^{\text{tot}}\rangle_{\tilde n_k}$
given by Eq.(\ref{NtotNk})
and $P(n^{\text{tot}}| \tilde n_k)$ given by Eq.(\ref{PntotOmega}) yields
\begin{equation}
\tilde \Delta_2^{\omega} = k^2\sigma_{\tilde n_k}^2 \frac{
1-2\,\omega\,\tilde n_k+
             \omega^2 [\tilde n_k^2-\sigma_{\tilde n_k}^2]}{ 
  (1-\omega\,\tilde n_k)^2} 
\,\,\approx \,\, k^2\sigma_{\tilde n_k}^2 [1-\omega^2\sigma_{\tilde n_k}^2]
\label{D2xOmega}
\end{equation}
Analytical expression for $\tilde \Delta_4^{\omega}$ 
can be found using Eq.(\ref{D4x}). Assuming $C_2^{\text{\,true}}\!=\,0$ one has
\begin{equation}
\tilde \Delta_4^{\omega} = 3\,k^4\sigma_{\tilde n_k}^4 \bigg[ 1 -
\frac{\sigma_{\tilde n_k}^4 \omega^4
    }{(1-\tilde n_k\,\omega)^4} -
\frac{2\,\sigma_{\tilde n_k}^2 \omega^2
    }{(1-\tilde n_k\,\omega)^2} \bigg]
\,\,\approx \,\, 3k^4\sigma_{\tilde n_k}^4 [ 1 - 2\,\omega^2\sigma_{\tilde n_k}^2]
\label{D4xOmega}
\end{equation}

If probability $P(n^{\text{tot}}) = const\,$  
correction term $\tilde \Delta_3\longrightarrow 0\,$
for $\tilde n_k$ large.
For $\bar\omega \neq 0$ which means $P(n^{\text{tot}}) \neq const\,$ 
correction $\tilde \Delta_3^{\omega}$
can be found using Eq.(\ref{D3x}):
\begin{equation}
\tilde \Delta_3^{\omega}=
-\frac{2\,k^3\omega^3\sigma_{\tilde n_k}^6}{
       (1-\tilde n_k \,\omega)^3 }\,\, \approx \,
-2\,k^3\omega^3\sigma_{\tilde n_k}^6  
\label{D3xOmega}
\end{equation}

For sample of events with constant $P(n^{\text{tot}})$
distribution (which means constant measured $P(\tilde n_k)$ distribution) 
one has $\omega \rightarrow 0$;\, 
$\langle n^{\text{tot}}\rangle_{\tilde n_k}=\tilde n_k \cdot \xi\,$;\, 
$\langle \tilde n_k \rangle_{\tilde n_k} = \tilde n_k$ and
$\tilde \Delta_n^{\omega}\rightarrow \tilde \Delta_n$.

Since 
acceptance parameter $\xi $ is not present in 
Eq.(\ref{D2xOmega},\ref{D4xOmega},\ref{D3xOmega})
fixed-multiplicity correction terms 
can be determined from experimentally accessible
quantities: $k^2$, $\omega^2$ and 
$\sigma_{\tilde n_k}^2$. 

\vspace{0.5cm}

\section{Relations between correlators}
Let us assume that mean transverse momentum $\bar p_t$ of particles in
events with fixed total multiplicity $n^{\text{tot}}$
(e.g. selected from the output of a MC event generator)
fluctuates around global mean
$\langle \bar p_t \rangle = \sum_{k=1}^{N_{ev}} \bar p_t^{\,i} /N_{ev}$ with probability
distribution
\begin{equation}
P(\bar p_t^{\,i}) = \frac{1}{\sqrt{2\pi\,}\,\sigma_{\bar p_t}} 
           e^{-(\bar p_t^{\,i} - \langle \bar p_t \rangle)^2/2\sigma^2_{\bar p_t}}
\label{Ppti}
\end{equation}
In this case $\sum_{k=1}^{N_{ev}} (\bar p_t^{\,i} -\langle \bar p_t\rangle)^2/N_{ev} 
               = \sigma^2_{\bar p_t}\,$ and one obtains
\begin{equation}
C_2^{\,\text{calc}} = C_2^{\,\text{true}} + \sigma^2_{\bar p_t}
\label{FFC2}
\end{equation}

Thus $C_2^{\,\text{calc}}$ correlator contains $\sigma^2_{\bar p_t}$ contribution
from event-by-event fluctuations of observable mean $\bar p_t$ and 
$C_2^{\,\text{true}}$ contribution 
from genuine two-particle correlations.
This can be verified directly using a suitable MC event generator. For
fluctuations of observable mean given by Eq.(\ref{Ppti})
one has $\sum_{k=1}^{N_{ev}} (\bar p_t^k -\langle \bar p_t\rangle)^4/N_{ev} 
               \approx 3\,\sigma^4_{\bar p_t}$ and in agreement with Eq.(\ref{C4sm})
\begin{equation}
C_4^{\,\text{calc}} = C_4^{\,\text{true}} + 3\,\sigma^4_{\bar p_t} 
                                          + 6\,\sigma^2_{\bar p_t} C_2^{\,\text{true}}
\label{FFC4}
\end{equation}
Assuming $C_4^{\,\text{true}}\!\rightarrow 0\,$
one can try to separate $\sigma^2_{\bar p_t}$ and $C_2^{\,\text{true}}$
contributions as 
\begin{equation}
(C_2^{\,\text{true}})^2 = (C_2^{\,\text{calc}})^2 - C_4^{\,\text{calc}}/3
\end{equation}
which is a solution of Eqs.(\ref{FFC2},\ref{FFC4}) 
for vanishing 4-particle correlations
($C_4^{\,\text{true}}\!= 0$).

\section{Conclusions}
A simple analytical calculation has shown that 
systematical shifts $\tilde \Delta_n$ in calculated values of correlators
are generated if
observable mean of the quantity under study is multiplicity-dependent
$\bar x(n_k)\!\neq\! const$. One can subtract
such systematical effects from the calculated correlators $C_n^{\,\text{calc}}$
to obtain "true" correlators using:\, 
$C_n^{\,\text{true}} = C_n^{\,\text{calc}}\! - \tilde \Delta_n$.

\section{APPENDIX}
Fixed-multiplicity correction term given by Eq.(\ref{DT2a})
contains quantity $\sigma_n(\xi,\tilde n_k)$ to be determined from MC simulation.  
We will show
that width $\sigma_n(\xi,\tilde n_k)$ of $n^{\text{tot}}_i$ fluctuations at given 
fixed measured $\tilde n_k^*$ can be
related to the width $\sigma_{\tilde n_k}$ of
$\tilde n_k$ fluctuations at fixed $n_*^{\text{tot}}\!\!\approx\!\xi\cdot \tilde n_k^*\,$ 
as $\,\sigma_n(\xi,\tilde n_k^*) = \xi \cdot \sigma_{\tilde n_k}(n_*^{\text{tot}})$ where 
$\xi $ is acceptance 
parameter $\xi = n^{\text{tot}}/\langle \tilde n_k \rangle $  and the width
$\sigma_{\tilde n_k}$ can be expressed as $\sigma_{\tilde n_k}(n_*^{\text{tot}}) = 
  \langle \tilde \sigma \rangle \sqrt{\tilde n_k^*}\,$.

To demonstrate this
let us divide given detector acceptance $\Omega_\xi $ into $N$
acceptance sub-regions $\omega_1 + \omega_2 + \ldots + \omega_N = \Omega_\xi$
in such a way that in every acceptance region $\omega_i$
equal average number of particles 
$\langle \tilde n^{\omega_i}\rangle =\langle\tilde n^{\omega_j}\rangle=\langle\tilde n_k\rangle/N$ 
will be measured.
Using $N = \tilde n_k\,$ acceptance subregions one has 
$\langle\tilde n^{\omega_i}\rangle \approx 1$ and 
$\sum_{i=1}^{N} \langle\tilde n^{\omega_i}\rangle = \langle \tilde n_k\rangle$.
For events with fixed total multiplicity $n^{\text{tot}}$ and
average measured multiplicity $\langle \tilde n_k\rangle$ the 
number of particles $\tilde n^{\omega_i}_k$ observed
in acceptance region $\omega_i$ 
will fluctuate (event-by-event) 
with some probability distribution $P_{\omega_i}(n^{\omega_i}_k)$ characterized
by the mean $\langle\tilde n^{\omega_i}\rangle $ and variance $\sigma_{\!\omega_i}\,$.

Assuming that probability distributions $P_{\omega_i}(n^{\omega_i}_k)$ satisfy
conditions for the applicability of generalized (e.g. m-dependent) Central Limit Theorem 
one can write
\begin{equation}
P_{\Omega_\xi}(\tilde n_k ) = 
P_{\Omega_\xi}\bigg(\sum_{i=1}^{N}\tilde n^{\omega_i}_k\bigg) \approx
\frac{e^{-(\tilde n_k - \langle \tilde n_k\rangle)^2/2\sigma^2_{\tilde n_{\!k}}} }
     {\sqrt{2\pi \sigma^2_{\tilde n_{k}}\,}} = P_{n^{\text{tot}}}(\tilde n_k)
= P(\tilde n_k|n^{\text{tot}})
\label{AXpnk}
\end{equation}
where $\sigma_{\tilde n_k} \approx
    \sqrt{\sigma_{\!\omega_1}^2 + \sigma_{\!\omega_2}^2 + \ldots +\sigma_{\!\omega_{N}}^2}
  = \sqrt {N\,}\langle \tilde \sigma \rangle\,$ (denoting
$(\sum_{i=1}^N \sigma^2_{\!\omega_i}/N)^{1/2}\! =\! \langle \tilde \sigma \rangle $). 
This suggests that for $\langle\tilde n_k\rangle =n^{\text{tot}}/\xi\,$ 
large enough the probability distribution 
of events with measured multiplicities $\tilde n_k$ in the set of events
with fixed total multiplicity $n^{\text{tot}}$ tends to be Gaussian 
and its width $\sigma_{\tilde n_k}$ increases with 
measured multiplicity as
\begin{equation}
\sigma_{\tilde n_k} = c\cdot \sqrt{\langle \tilde n_k\rangle} 
\label{X51}
\end{equation}

One can ask a similar question in the other way around: 
What are the fluctuations
of total multiplicity of particles $n_i^{\text{tot}}$ for events with
fixed observed multiplicity $\tilde n_k$~?

Using the Bayes' theorem \cite{Papoulis} one can
calculate probability
$P(n^{\text{tot}} | \tilde n_k)$
of observing the event with total multiplicity $n^{\text{tot}}$ in the
subset of events with fixed measured multiplicity $\tilde n_k$ 
\begin{equation}
P(n^{\text{tot}} | \tilde n_k) = \frac{
   P(\tilde n_k | n^{\text{tot}}) P (n^{\text{tot}})}{
   \int P(\tilde n_k | n^{\text{tot}}) P(n^{\text{tot}})\,dn^{\text{tot}} }
\label{BayesSS}
\end{equation}
where $P(A|B)$ denotes a conditional probability of observing 
quantity $A$ for given $B$
and $P (n^{\text{tot}})$ is the prior probability\footnote{Note, that
$P (n^{\text{tot}})$ depends on a particular setting of the detector trigger.} 
of event with 
total multiplicity $n^{\text{tot}}\!$. 
Choosing the sample of events with $P(n^{\text{tot}}) = const\,$ 
simplifies
the situation and one has 
$P(n^{\text{tot}} | \tilde n_k) = \lambda\cdot P(\tilde n_k | n^{\text{tot}})$
where $\lambda$ is a normalization constant. 

For $\langle \tilde n_k \rangle$ large enough 
$P(\tilde n_k|n^{\text{tot}})$ 
given by Eq.(\ref{AXpnk}) can be used to obtain 
probability $P(n^{\text{tot}}|\tilde n_k)$ of having
event with total multiplicity $n^{\text{tot}}$ in the group of events
with fixed measured multiplicity $\tilde n_k$. For event sample
with probability $P(n^{\text{tot}}) = const\,$ one has:
\begin{equation}
P(n^{\text{tot}}|\tilde n_k) = \lambda \cdot P(\tilde n_k | n^{\text{tot}})
= \lambda\frac{
e^{-(\xi\cdot\tilde n_k - \xi \langle \tilde n_k\rangle)^2/2
\xi^2\sigma^2_{\tilde n_{\!k}}} }
     {\sqrt{2\pi \sigma^2_{\tilde n_{k}}\,}} 
= \frac{
e^{-(n^{\text{tot}} - \tilde n_k\cdot \xi)^2/2\sigma^2_{\text{tot}} } }{ 
      \sqrt{2\pi \sigma^2_{\text{tot}}\, } }
\label{AXpntot}
\end{equation}
where we have denoted $\xi\cdot \sigma_{\tilde n_k} = \sigma_{\text{tot}}$,
used $n^{\text{tot}}/\xi = \langle \tilde n_k\rangle$ and normalized
the resulting Gaussian distribution to unity. 
For event sample with 
probability $P(n^{\text{tot}})=2\bar\omega(1-\bar\omega \, n^{\text{tot}})$
evaluation of denominator in Eq.(\ref{BayesSS}) gives\,
$2\,\xi\,\bar \omega (1-\xi\, \tilde n_k\, \bar\omega)$ and
Eq.(\ref{BayesSS}) then yields 
\begin{equation}
P(n^{\text{tot}}|\tilde n_k)
=\frac{e^{-(n^{\text{tot}} -\tilde n_k\cdot \xi)^2/2\xi^2\sigma^2_{\tilde n_k}} }{
       \sigma_{\tilde n_k}\sqrt{2\pi\,}\,\xi } 
 \frac{(1-n^{\text{tot}} \bar\omega )}{
       (1-\tilde n_k\xi\, \bar\omega )}
\label{PntotOmegaX}
\end{equation}
which is valid for fluctuations of $n^{\text{tot}}$
(at given fixed $\tilde n_k$)
within the range of $P(n^{\text{tot}})$
distribution. For $P(n^{\text{tot}})=2\bar\omega(1-\bar\omega \, n^{\text{tot}})$ 
this means
$\tilde n_k\xi + 3\sigma_{\text{tot}} < 1/\bar\omega\,$ which
keeps $\,\xi\,\tilde n_k\, \bar\omega < 1$.
 
For Poissonian
probability $P(\tilde n_k) =
\langle \tilde n_k \rangle ^{\tilde n_k}\cdot e^{-\langle\tilde n_k\rangle}/ 
\tilde n_k !\,\,$ of measuring
$\tilde n_k$ particles in a detector (exposed to events with fixed
multiplicity $n^{\text{tot}}\!\!=\! \xi\langle \tilde n_k \rangle$)
Bayes' law 
gives asymmetrical probability of $n^{\text{tot}}$ for a given fixed $\tilde n_k$:
\begin{equation}
\tilde {\cal{P}}(n^{\text{tot}} | \tilde n_k ) = \lambda \frac {
(n^{\text{tot}}/\xi)^{\tilde n_k}\cdot e^{-n_{\text{tot}}/\xi} } {
\tilde n_k ! } P(n^{\text{tot}}) = 
\lambda' \cdot e^{-n^{\text{tot}}/\xi}\cdot (n^{\text{tot}})^{\tilde n_k}
\cdot P(n^{\text{tot}})
\label{FPX}
\end{equation}
(here $\lambda$ and $\lambda'$ are normalization factors). 
Fluctuations of $n^{\text{tot}}$ at small fixed $\tilde n_k$
can be approximated by this function  
(see Fig.3 for 
$\tilde {\cal{P}}(n^{\text{tot}} | \tilde n_k )$ at $\tilde n_k = 10$ 
using $P(n^{\text{tot}})=const$).  

\vspace{0.5cm}

{\bf{Acknowledgements:}} 
The author is grateful to R.Lednicky, J.Manjavidze and S.Shimanskii for
illuminating discussions, comments and motivation and to
Laboratory of Particle Physics at JINR
 in Dubna for kind hospitality.
This work was supported also by Slovak Grant Agency for Sciences VEGA under
grant N. 2/7116/27.

\end{document}